\documentclass[aps,pra,reprint,showpacs,floatfix,preprintnumbers]{revtex4-1}

\usepackage{graphicx}
\usepackage{color}
\usepackage{amssymb}
\usepackage{amsmath}
\usepackage{epsfig}
\usepackage{bm}
\usepackage[american]{babel}
\usepackage{xfrac} 
\usepackage[colorlinks,urlcolor=blue,citecolor=blue,linkcolor=blue,pdfstartview=FitH]{hyperref}
\usepackage[T1]{fontenc} 
\usepackage{txfonts} 

\DeclareGraphicsExtensions{.pdf,.png,.jpg}
\newcommand{\interval}{I}
\newcommand{\relangle}{\theta_{21}}
\newcommand{\Lcrit}{L_\mathrm{cr}}

\begin{document}
\title{Hamiltonian dynamics of two same-sign point vortices}
\author{Anderson V. Murray}
\author{Andrew J. Groszek}
\author{Pekko Kuopanportti}
\author{Tapio Simula}
\affiliation{School of Physics and Astronomy, Monash University, Victoria  3800, Australia}

\date{March 29, 2016}
\begin{abstract}
We have studied numerically the Hamiltonian dynamics of two same-sign point vortices in an effectively two-dimensional, harmonically trapped Bose--Einstein condensate. We have found  in the phase space of the system an impenetrable wall that divides the dynamics into two distinct and exhaustive types. In the two-dimensional position-coordinate space, the first type corresponds to intersecting single-vortex orbits and the second type to orbits that have no points in common. The two types are also easily distinguished in the two-dimensional space spanned by the radial and angular velocities of the vortices: In the first type, both single-vortex orbits are the same simple loop in this two-dimensional space, whereas in the second type the two orbits constitute two nonintersecting loops. The phase-space-dividing wall is distinct from the bifurcation curve of rigidly rotating states found by Navarro \emph{et al.} [\href{http://journals.aps.org/prl/abstract/10.1103/PhysRevLett.110.225301}{Phys. Rev. Lett. {\bf 110}, 225301 (2013)}]. 
\end{abstract}
\preprint{DOI: \href{http://dx.doi.org/10.1103/PhysRevA.93.033649}{10.1103/PhysRevA.93.033649}}
\maketitle

\section{Introduction}\label{sec:intro}
The mathematical underpinnings of the dynamics of pointlike vortices in classical fluids were established in the nineteenth century~\cite{Helmholtz1867a,Kirchhoff1876a,Routh1880a,Lin1941a,Lin1941b}. In the 1940s, Lars Onsager realized that the equations of motion describing such vortices in two dimensions are mathematically equivalent to Hamilton's equations of motion for particles moving in one spatial dimension and that a large collection of them could be treated with the machinery of statistical mechanics~\cite{Onsager1949a,Eyink2006a}. His motivation was to develop understanding of fluid turbulence by describing the statistical properties of the turbulent fluid in terms of a collection of point-vortex particles. He predicted that in turbulent two-dimensional (2D) fluid flows, the vortices, rather than being randomly distributed throughout the fluid, should arrange into giant clusters of the order of the size of the system~\cite{Onsager1949a}. Although such \emph{Onsager vortices} appear to be prevalent in many classical fluid flows ranging from large-scale ocean currents and planetary atmospheric flows~\cite{Mil1992.PRA45.2328} to thin liquid films~\cite{Tab1991.PRL67.3772,Par1998.PhysFluids10.3126}, in the classical context the point-vortex model is often dismissed as an oversimplified description of real fluids. 

In the context of quantum mechanics, vortex-filament models in three-dimensional (3D) systems~\cite{Schwarz1985a} and the corresponding point-vortex models in two dimensions have been applied extensively to superfluid helium, where the quantization of circulation provides justification for treating them within the point-vortex approximation, as already noted by Onsager~\cite{Onsager1949a}. However, quantitative comparisons of vortex dynamics with experiments have been difficult to achieve due to the considerable challenges of imaging individual vortex lines with\ {\AA}ngstr\"om-scale core sizes~\cite{Bewley2006a,Fonda2014a}. A renewed interest in the point-vortex models~\cite{Pointin1976a,Aref1983a,Campbell1991a,Viecelli1995a} has emerged in superfluids of ultracold atomic gases~\cite{Freilich2010a,Kuopanportti2011a,Middelkamp2011a,Navarro2013a,Moo2015.PRA92.051601,Torres2011a,Torres2011b} where the vortex cores can be sufficiently large to be resolved optically even \emph{in situ}~\cite{Wilson2015a} and their circulation direction might be detected using, e.g., the vortex gyroscope imaging technique~\cite{Powis2014a}. 

Recent experiments~\cite{Freilich2010a,Kuopanportti2011a,Middelkamp2011a,Navarro2013a,Moo2015.PRA92.051601} have shown that the vortices in the superfluid gases are amenable to the point-vortex approach, opening up further possibilities for quantitative studies of vortex dynamics such as Kelvin waves~\cite{Bretin2003a,Fetter2004a,Simula2008b,Simula2008a,Koens2013a}, Crow instabilities~\cite{Crow1970a,Berloff2001a,Simula2011a}, and Tkachenko vortex waves~\cite{Tkachenko1966a,Tkachenko1966b,Sonin1987a,Coddington2003a,Baym2003a,Simula2004a,Simula2005a,Simula2010a,Simula2011a,Simula2013a,Simula2013b}. 
Understanding such few-vortex phenomena forms the basis for solving more complex problems involving vortices; a topical example is quantum turbulence in 2D systems~\cite{Kobayashi2005a,Parker2005a,Bradley2012a,Neely2013a,Reeves2013a,Kwon2014a,Reeves2014a,Billam2014a,Simula2014a,Stagg2015a,Billam2015a} and the emergence of Onsager vortices and negative Boltzmann temperatures for vortices in disk-shaped traps~\cite{Simula2014a,Groszek2015a}. 

In 2D superfluids, strong turbulence is tantamount to chaotic dynamics of the quantized vortices in the system. For three or more vortices in an effectively 2D Bose--Einstein condensate (BEC) confined by a cylindrically symmetric, highly oblate harmonic trap, the vortex dynamics can become chaotic. However, a two-vortex problem is integrable due to two conservation laws related to the energy and angular momentum of the system. As a precursor to studying the onset of turbulence, we focus here on the problem of two vortices of the same circulation. Our work is motivated by recent BEC experiments that discovered a bifurcation of rigidly rotating stationary states in the two-vortex case~\cite{Navarro2013a}. Here we find that the phase space of the two-vortex system is divided into two topologically distinct regions corresponding to two radically different types of two-vortex motion: In one region the individual orbits of the two vortices overlap, whereas in the other region the orbits never cross each other. 

\begin{figure*}[h!t]
\includegraphics[width=0.9\textwidth,keepaspectratio]{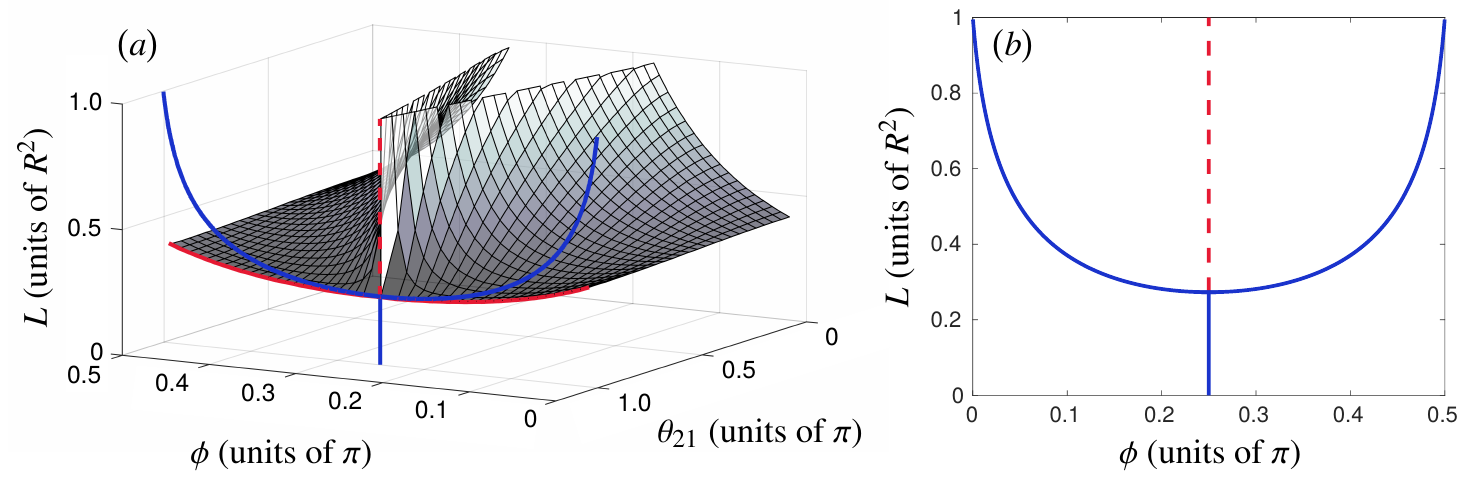}
\caption{\label{fig:schematic} Phase-space-dividing wall (a) and the curve of rigidly rotating states (a) and (b). In (a), the full three-dimensional parameter space $\left(\phi,L,\relangle \right)$ of possible two-vortex configurations is shown. Above the surface, the individual orbits of the two vortices never cross each other, whereas below it the orbits intersect. Here $\phi=\textrm{tan}^{-1}\left(r_2 /r_1\right)$, $L = r_1^2 + r_2^2$, and $\relangle=\theta_2-\theta_1$, with $\{ (r_k,\theta_k )\}$ denoting the polar coordinates of the vortices. The (blue) solid curve corresponding to rigidly rotating states lies in the plane $\relangle=\pi$ shown in (b); see also Fig.~\ref{fig:phase_transitions}. The (red) dashed line marks the unstable rigidly rotating states occurring for $L>\Lcrit\approx 0.273\,R^2$. Panel (b) corresponds to Fig.~1(c) in Ref.~\cite{Navarro2013a}. Notice that the rigidly rotating states trace a one-dimensional curve in the three-dimensional parameter space (a), whereas the red curves are the boundary curves of the two-dimensional phase-space-dividing wall for $\relangle=\pi$. As in Ref.~\cite{Navarro2013a}, all results are for $\Omega_\mathrm{int}/\Omega_0=0.1$~[Eqs.~\eqref{eq:PVM}].} 
\end{figure*}

Figure~\ref{fig:schematic} summarizes the main findings. The wall that divides the two-vortex phase space is shown in Fig.~\ref{fig:schematic}(a), indicating a sharp transition between the two types of motion as the initial vortex positions are varied. This transition boundary is impenetrable in the sense that any two-vortex state not located on the boundary at any one time will always remain at that side of the boundary as the system evolves in time. Furthermore, we find that the asymmetric rigidly rotating states investigated in Ref.~\cite{Navarro2013a} and shown here in Fig.~\ref{fig:schematic}(b) are not related to this change in the topology of the accessible phase space.

The remainder of this article is organized as follows. In Sec.~\ref{sec:methods}, we outline the point-vortex model we use for describing the effectively 2D dynamics of vortices in a highly oblate harmonically trapped BEC. Section~\ref{sec:results} presents our results using two complementary descriptions: the position-space representation and the velocity-space representation. These are used for revealing the phase-space boundary that separates the system dynamics into two topologically distinct classes. The paper is concluded with a discussion in Sec.~\ref{sec:discussion}. 

\section{Point-vortex model}\label{sec:methods}

We restrict our attention to effectively 2D dynamics where quantized vortices in a harmonically trapped BEC can be modeled as point particles moving in the $xy$ plane. We take the velocity of each vortex to be the sum of two independent contributions: (i)~solitary orbital motion due to an axisymmetric harmonic trap, and (ii)~motion in the flow field of all other vortices. Thus, we consider the following system of $N$ coupled ordinary differential equations~\cite{Middelkamp2011a,Navarro2013a,Moo2015.PRA92.051601,Torres2011a}:
\begin{equation}\label{eq:PVM}
-i \dot{z}_k = R^2 \Omega_0  \frac{s_k z_k}{R^2 -|z_k|^2} + R^2  \Omega_\mathrm{int}  \sum\limits_{l\,\left(\neq k\right)}^N s_l \frac{z_k-z_l}{|z_k-z_l|^2},
\end{equation}
where $z_k = x_k + i y_k$, $\left(x_k,y_k\right)$ are the position coordinates of the $k$th vortex, $s_k \in \mathbb{Z}$ is its circulation number, and $k\in\left\{ 1,\dots,N\right\}$. The vortices are constrained to move within the Thomas--Fermi radius of the condensate, $R > |z_k|$. Furthermore, $\Omega_0$ is the orbital angular frequency of a solitary unit-strength vortex infinitesimally close to the trap center~\cite{Freilich2010a,Fetter2001a}, and $\Omega_\mathrm{int}$ is an angular frequency determining the effective strength of the vortex--vortex interaction~\cite{Navarro2013a}.

Equations~\eqref{eq:PVM} are equivalent to Hamilton's equations of motion,
\begin{equation} \label{eq:HamiltonsEq}
\begin{aligned}
s_k \dot{x}_k & = \frac{\partial H}{\partial y_k}, \\
s_k \dot{y}_k & = - \frac{\partial H}{\partial x_k},
\end{aligned}
\end{equation}
corresponding to the Hamiltonian
\begin{equation} \label{eq:PVMHamiltonian}
\begin{aligned}
H & = \frac{R^2\Omega_0}{2} \sum\limits_{k=1}^N s_k^2 \textrm{ln}\left(1 - \frac{|z_k|^2}{R^2} \right)\\&- R^2 \Omega_\mathrm{int} \sum\limits_{k=1}^N \sum\limits_{l\,\left(> k\right)}^N s_k s_l \textrm{ln} \frac{|z_k - z_l|}{R}.
\end{aligned}
\end{equation}
Therefore, the point-vortex model has the peculiar feature that the coordinate space $\{(x_k,y_k)\}$ can also be interpreted as the Hamiltonian phase space of the system. In addition to $H$, the model also has another integral of motion, $L=\sum_k s_k |z_k|^2$, due to the underlying rotational symmetry. In analogy to point particles, this quantity is referred to as the \emph{point-vortex angular momentum}; note, however, that it should not be confused with the orbital angular momentum that the vortex induces in the flow of the surrounding superfluid. In fact, whereas the point-vortex angular momentum $s_k |z_k|^2$ of a single vortex increases as the vortex moves away from the symmetry axis, the angular momentum of the superfluid decreases under such circumstances.
 
By denoting $z_k = r_k \exp\left(i \theta_k\right)$, we obtain the radial and angular vortex velocities $\dot{r}_k$ and $\dot{\theta}_k$, respectively, from the Cartesian velocities as
\begin{equation} \label{eq:PolarVel}
\left[ \begin{array}{c} \dot{r}_k \\ r_k \dot{\theta}_k \end{array} \right] =
\begin{bmatrix} \textrm{cos}\theta_k & \textrm{sin}\theta_k \\ -\textrm{sin}\theta_k & \textrm{cos}\theta_k \end{bmatrix}
\left[ \begin{array}{c} \dot{x}_k \\ \dot{y}_k \end{array} \right].
\end{equation}
The velocity space $\{ ( \dot{r}_k,\dot{\theta}_k ) \}$ turns out the be extremely useful for representing the vortex dynamics in subsequent analysis (Sec.~\ref{subsec:velocity_space}).

When is the point-vortex model applicable? In 3D BECs, vortex filaments can be described as point particles as long as they remain straight and parallel to one another, rendering their dynamics effectively 2D. Hence, the validity of  Eqs.~\eqref{eq:PVM}--\eqref{eq:PolarVel} extends beyond the regime of quasi-2D BECs confined extremely tightly in the $z$ direction. For instance, although the BECs of Refs.~\cite{Freilich2010a,Middelkamp2011a,Navarro2013a} were 3D, very good agreement with the point-vortex description was obtained, likely because the trapping along the $z$ axis was strong enough to limit vortex motion to the $xy$ plane and suppress vortex bending and tilting away from the $z$ direction. Aspects of condensate dimensionality in regards to vortices and Kelvin waves were studied theoretically in Ref.~\cite{Roo2011.PRA84.023637}, further indicating that sufficiently oblate yet still 3D BECs may be considered 2D as far as vortex dynamics and quantum turbulence are concerned. 

From here on, we focus on a system of two vortices with equal circulations, setting $N=2$ and $s_1=s_2=1$. We measure lengths in units of $R$ and time in units of $\Omega_0^{-1}$. Up to a rotation of the coordinate system, all possible two-vortex configurations are spanned by three variables: the angle $\phi=\textrm{tan}^{-1}\left(r_2/r_1\right)$, the point-vortex angular momentum $L = r_1^2 + r_2^2$, and the azimuthal angle $\relangle=\theta_2-\theta_1$ between the two vortices. Recently, Navarro \emph{et al.}~\cite{Navarro2013a} investigated this system both theoretically and experimentally for two-vortex configurations with $\relangle=\pi$. They demonstrated that when $\Omega_\mathrm{int}/\Omega_0=0.1$ (a value that we adopt throughout this work), the system exhibits a pitchfork bifurcation at $L=\Lcrit\approx 0.273\,R^2$ that induces the emergence and stabilization of asymmetric ($r_1 \neq r_2$) rigidly rotating vortex configurations and the destabilization of symmetric ($r_1=r_2$, i.e., $\phi=\pi/4$) rigidly rotating states at $L>\Lcrit$.

We have solved Eqs.~\eqref{eq:PVM} numerically using the ode113 function in {\footnotesize MATLAB} with a relative tolerance of $10^{-13}$, absolute tolerance of $10^{-15}$, and a variable time step.
As the initial conditions $\left(\phi,L,\relangle \right)$, we consider 20 equidistant values from $\phi=0.238\,\pi$ to $\phi=0.466\,\pi$ and from $L=0.1\,R^2$ to $L=0.955\,R^2$, and 10 equidistant values from $\relangle=0.1\,\pi$ to $\relangle=\pi$. For trajectories that have exactly symmetric initial conditions ($\phi=\pi/4$ and $\relangle=\pi$) and exhibit stable rigid-body rotation ($L<\Lcrit$), our simulations show deviations from the initial radius $r_k\left(0\right)$, initial Hamiltonian energy $H$, and initial $L$ of at most $10^{-6}$ in the respective units of each over time intervals under consideration.

\section{Results}\label{sec:results}

In this section, we present our numerical results on the dynamics of two same-sign point vortices and, in particular, describe the emergence of the two distinct classes of motion in the system.  To relate our results to the findings of Ref.~\cite{Navarro2013a}, we limit the specific examples to the case $\relangle=\pi$ corresponding to vortices that are initially located on opposite sides of the center of the harmonic trap. However, we emphasize that the two distinct phase-space regions persist for all values of $\relangle$ [Fig.~\ref{fig:schematic}(a)]. We first consider the position-coordinate representation (Sec.~\ref{subsec:position_space}) in order to provide a physically intuitive picture but subsequently switch to using the radial and angular velocities as our coordinates (Sec.~\ref{subsec:velocity_space}) because the emergence of the two types of motion is most apparent in this representation.

\subsection{Position-space representation}\label{subsec:position_space}

\begin{figure}[tb]
\includegraphics[width=\columnwidth,keepaspectratio]{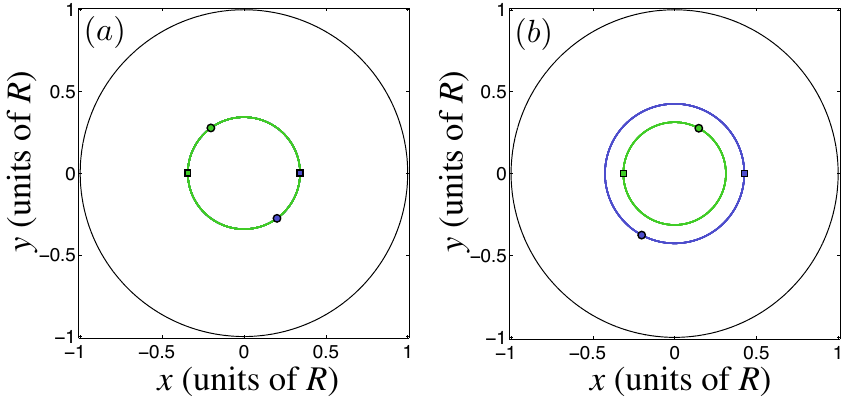}
\caption{\label{fig:rigid_rot_stable} Dynamics of rigidly rotating configurations of two same-sign point vortices. (a) Symmetric rigidly rotating configuration with the initial conditions $\phi / \pi = 0.25$, $L = 0.235R^2$, and $\relangle=\pi$, corresponding to panel B17 in Fig.~\ref{fig:phase_transitions}. In this and all other figures, the total simulation time is $60\Omega_0^{-1}$. (b) Asymmetric rigidly rotating state with $L = 0.28R^2$, $\phi / \pi = 0.298$, and $\relangle=\pi$ (panel F16). The initial and final position of each vortex are denoted by square and circular markers, respectively. The orbit of vortex 1 is shown in dark (blue) color and that of vortex 2 in light (green) color. In panel (a), the individual orbits of the two vortices are the same.} 
\end{figure}

Consider first two same-sign vortices placed at equal distances on opposite sides of the trap center, i.e., $\phi=\pi/4$ and $\relangle=\pi$, in terms of their position coordinates $\left(x_k,y_k\right)\in\mathbb{R}^2$, $r_k < R$. These states lie on the solid vertical line segment in Fig.~\ref{fig:schematic}(b). As long as $L < \Lcrit \approx 0.273\,R^2$~\cite{Navarro2013a}, the resulting motion will consist of stable rigid-body rotation as exemplified in Fig.~\ref{fig:rigid_rot_stable}(a). The dynamics of this state show no major divergence from rigid rotation over time scales of $\sim 4000\,\Omega_0^{-1}$ and satisfy $|r_k(t)-r_k(0)|/R < 10^{-10}$ during the entire simulation. 

\begin{figure}[tb]
\includegraphics[width=0.75\columnwidth,keepaspectratio]{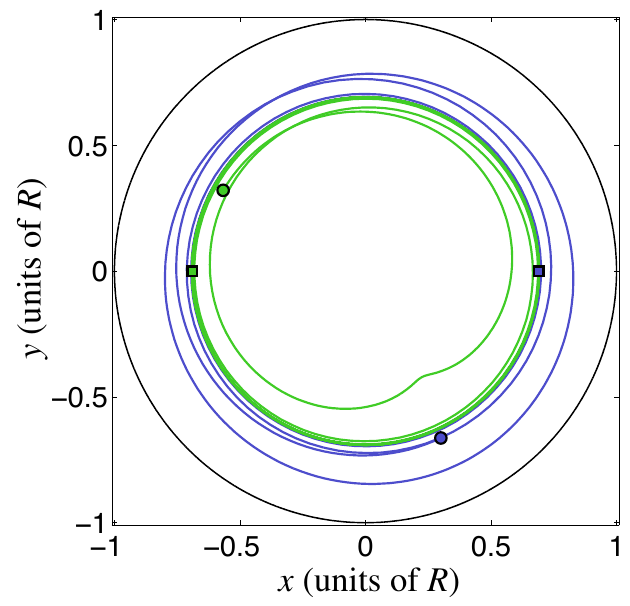}
\caption{\label{fig:symmetric_rigid_unstable} Simulated dynamics of an unstable symmetric rigidly rotating two-vortex configuration with the initial conditions $\phi/\pi = 0.25$, $L = 0.955R^2$, and $\relangle=\pi$, corresponding to panel B1 in Fig.~\ref{fig:phase_transitions}. Although in theory the configuration rotates rigidly, the instability causes even the smallest numerical errors to result in large deviations from the rigid rotation.} 
\end{figure}

On the other hand, it was recently found by Navarro \emph{et al.}~\cite{Navarro2013a} that when $L > \Lcrit$, the symmetric rigidly rotating states with $\relangle=\pi$ and $\phi=\pi/4$ are dynamically unstable due to a symmetry-breaking pitchfork bifurcation, and stable rigid-body rotation is instead exhibited by \emph{asymmetric} states with $\relangle=\pi$ and $\phi = \pi/4 \pm \delta$, where the specific value of $\delta$ is determined by $L$. In Fig.~\ref{fig:schematic}(b), the stable rigidly rotating two-vortex states lie on the solid curve, whereas the unstable symmetric rigidly rotating states are indicated by the dashed line segment. An example of an asymmetric rigidly rotating state is shown in Fig.~\ref{fig:rigid_rot_stable}(b). Figure~\ref{fig:symmetric_rigid_unstable}, in turn, illustrates the destabilization of the symmetric configuration for $L>\Lcrit$: The initial configuration is perfectly symmetric, but after a sufficiently long simulation time, the state becomes nonrigidly rotating since even the smallest numerical deviation pushes the vortices out of the rigidly rotating trajectories. 

\begin{figure}[tb]
\includegraphics[width=\columnwidth,keepaspectratio]{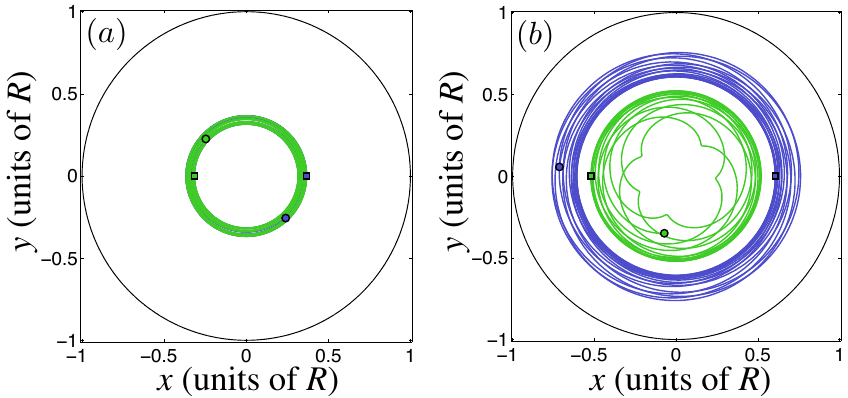}
\caption{\label{fig:general_motion_pos} (a) Two-vortex dynamics with overlapping orbits and initial conditions $\phi / \pi = 0.274$, $L = 0.235R^2$, and $\relangle=\pi$ (panel D17 in Fig.~\ref{fig:phase_transitions}). (b) Dynamics for which the vortex orbits never intersect; here $\phi / \pi = 0.274$, $L = 0.64\,R^2$, and $\relangle=\pi$ (panel D8).} 
\end{figure}

Next, we turn to the general case of two-vortex dynamics with any $\phi$, $L$, and $\relangle$, considering the full 3D configuration space [Fig.~\ref{fig:schematic}(a)]. Two possible types of general stable dynamics in the nonrigidly rotating configurations are observed. Figure~\ref{fig:general_motion_pos}(a) shows the first type, in which the vortices trace out orbits that are confined to the same spatial region of the trap and intersect each other at different times. If we define the closed intervals $\interval_k := \left[ \min_t r_k\left(t\right), \max_t r_k\left(t\right) \right]$, which describe the smallest annuli inside which each vortex moves, the first type of motion is characterized by $\interval_1=\interval_2$. Figure~\ref{fig:general_motion_pos}(b), in turn, is an example of the other general type of dynamics, in which the two vortices are confined to separate spatial regions and their orbits never intersect. In this case, $\interval_1 \cap \interval_2 = \emptyset$. The equivalence of the coordinate space $\{ (x_k,y_k ) \}$ to the Hamiltonian phase space of the system [Eq.~\eqref{eq:HamiltonsEq}] suggests that this difference between shared and separate trap regions represents a change in the topology of the system's accessible phase space. 

The mixing of two time scales due to the orbital and relative motion of the vortices makes it difficult to quantify the periodic motion of the vortices. To elucidate the relative motion of the vortices, we can transform to a rotating frame of reference. In this frame, the coordinate axes $x'$ and $y'$ are rotating relative to the laboratory frame with the time-dependent angular velocity $\big(\dot{\theta}_1+\dot{\theta}_2\big)/2$, i.e., the instantaneous average angular velocity of the two vortices.

\begin{figure}[tb]
\includegraphics[width=\columnwidth,keepaspectratio]{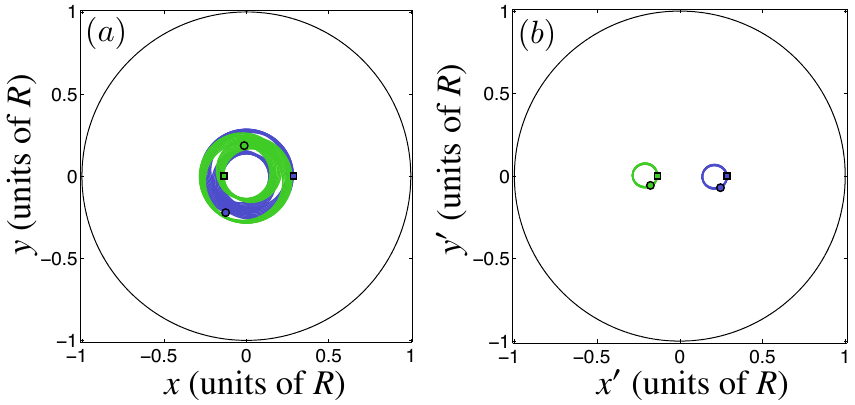}
\caption{\label{fig:2v_shared_rot_com} Comparison of laboratory- and rotating-frame perspectives with the initial conditions $\phi / \pi = 0.358$,  $L = 0.1\,R^2$, and $\relangle=\pi$, corresponding to panel K20 in Fig.~\ref{fig:phase_transitions}. (a) Laboratory-frame representation showing intersecting single-vortex orbits. (b) Rotating-frame view of the same dynamics showing orbits that are the same shape but at opposite sides of the trap. The coordinate axes $x'$ and $y'$ rotate with the instantaneous average angular velocity of the two vortices.} 
\end{figure}

\begin{figure}[tb]
\includegraphics[width=\columnwidth,keepaspectratio]{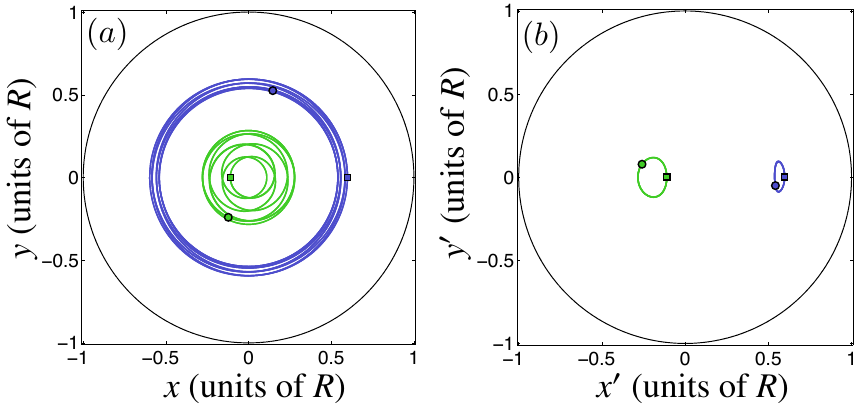}
\caption{\label{fig:2v_split_rot_com} Comparison of laboratory- and rotating-frame perspectives for the initial conditions $\phi / \pi = 0.442$, $L = 0.37\,R^2$, and $\relangle=\pi$, corresponding to panel R14 in Fig.~\ref{fig:phase_transitions}. (a) Laboratory-frame representation showing vortices moving in separate regions of the trap. (b) Rotating-frame view of the same dynamics showing orbits that are of different shape and at different distances from the trap center.} 
\end{figure}

The fixed and rotating frames of reference are compared for the case of intersecting orbits in Fig.~\ref{fig:2v_shared_rot_com} and for noncrossing orbits in Fig.~\ref{fig:2v_split_rot_com} (here again both examples start with $\theta_{12}=\pi$). When the orbits cross in the laboratory frame [Fig.~\ref{fig:2v_shared_rot_com}(a)], they form similarly shaped closed curves in the rotating frame [Fig.~\ref{fig:2v_shared_rot_com}(b)], which are centered at equal distances but at opposite sides of the trap center. For noncrossing orbits in the laboratory frame [Fig.~\ref{fig:2v_split_rot_com}(a)], the rotating frame yields two closed curves that have different shapes and are located at different distances from the trap center [Fig.~\ref{fig:2v_split_rot_com}(b)]. We conclude that whereas the overall vortex motion always reduces to relatively simple orbits in the rotating frame~\cite{Middelkamp2011a}, distinguishing between the two general classes of dynamics is not particularly simple. In addition, the small numerical errors in determining the correct frame-rotation frequency are prone to accumulate for long simulation times, leading to deviations from the simple closed curves.

\subsection{Velocity-space representation}\label{subsec:velocity_space}

\begin{figure}[tb]
\includegraphics[width=\columnwidth,keepaspectratio]{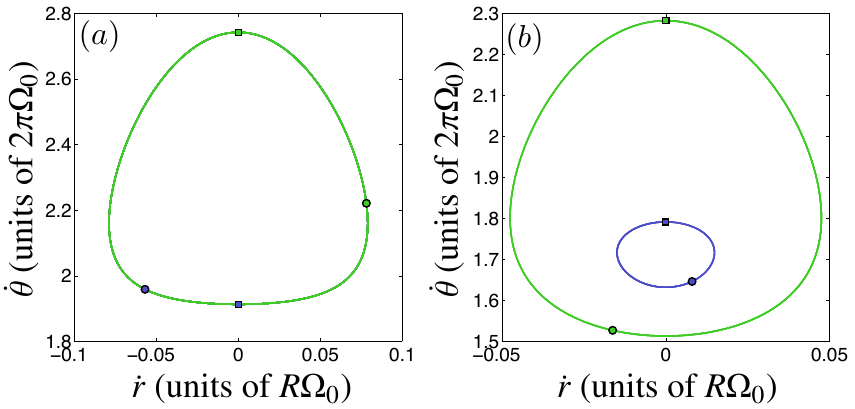}
\caption{\label{fig:polar_general} Two-vortex dynamics in the velocity-space representation. (a) Overlaid orbits showing that each vortex has exactly the same dynamic, albeit out of phase from the other. The initial conditions are $\phi / \pi = 0.358$, $L = 0.1R^2$, and $\relangle=\pi$, corresponding to Fig.~\ref{fig:2v_shared_rot_com} (panel K20 in Fig.~\ref{fig:phase_transitions}). (b) Vortex orbits for the initial conditions $\phi / \pi = 0.442$, $L = 0.37R^2$, and $\relangle=\pi$ (Fig.~\ref{fig:2v_split_rot_com} and panel R14 in Fig.~\ref{fig:phase_transitions}) showing that 
the two vortices trace separate loops in the polar velocity space.} 
\end{figure}

The two types of dynamics of the two-vortex system become particularly evident when one inspects the motion in terms of the radial and angular velocities $\{ ( \dot{r}_k,\dot{\theta}_k ) \}$ [Eq.~\eqref{eq:PolarVel}]. This method is invariant under the rotation of the vortex configuration about the trap center, and we refer to it as the  \emph{velocity-space representation}.  The two general types of two-vortex dynamics are illustrated using this representation in Fig.~\ref{fig:polar_general}. Figure~\ref{fig:polar_general}(a) shows the orbits that the vortices trace out in the 2D velocity space $( \dot{r},\dot{\theta})$ in the case where their individual real-space orbits intersect and $\interval_1 = \interval_2$. We observe that in this case both vortices always trace identical simple loops in the velocity space (for stable symmetric rigidly rotating states this loop contracts into a single point). Since the conservation of $H$ and $L$ guarantee that $(\dot{r}_1,\dot{\theta}_1) \neq (\dot{r}_2,\dot{\theta}_2)$ whenever $\dot{r}_k\neq 0$, the vortices traverse the joint velocity-space loop out of phase. The other type of general two-vortex motion, where their coordinate-space orbits never cross and $\interval_1\cap\interval_2=\emptyset$, is illustrated in the velocity space in Fig.~\ref{fig:polar_general}(b). In this case, the two vortices trace separate loops in the velocity space that do not intersect each other.

Let us next consider in detail what happens in the velocity-space when one crosses over from one type of motion to the other, i.e., crosses over the separating boundary in the initial configuration space $\left( L, \phi, \relangle\right)$  [Fig.~\ref{fig:schematic}(a)]. We stress that such a crossover can never occur during the dynamics; instead, one should think of varying the parameters $\left(\phi,L,\theta_{21}\right)$ manually. Again, we consider the case $\relangle=\pi$, due to its relevance to Ref.~\cite{Navarro2013a}.

\begin{figure}[tb]
\includegraphics[width=\columnwidth,keepaspectratio]{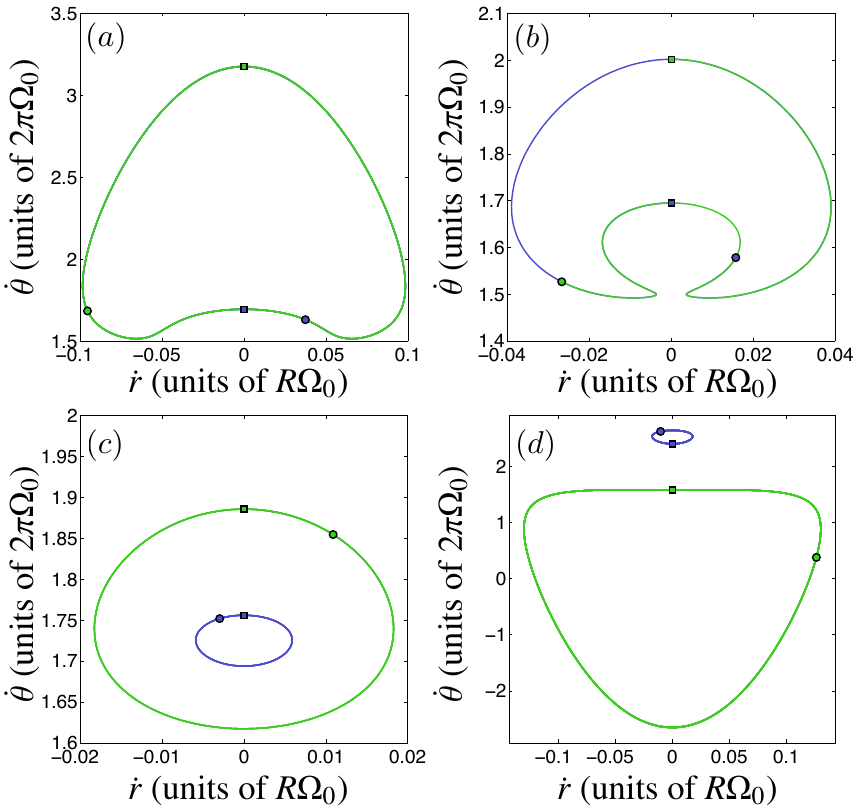}
\caption{\label{fig:polar_progression} Comparison of different types of observed dynamics in the polar velocity space. (a) Shared-space dynamics far from the phase transition with the initial conditions $\phi / \pi = 0.454$ and $L = 0.28\,R^2$ (panel S16 in Fig.~\ref{fig:phase_transitions}). (b) Shared-space dynamics near the transition with $\phi / \pi = 0.418$ and $L = 0.325\,R^2$ (panel P15). (c) Noncrossing dynamics near the transition with $\phi / \pi = 0.418$ and $L = 0.37\,R^2$ (panel P14). (d) Noncrossing dynamics far from the transition with  $\phi / \pi = 0.418$ and $L = 0.595\,R^2$ (panel P9). In each panel, the initial separation angle is $\relangle=\pi$.} 
\end{figure}

At sufficiently low $L$ values, the motion corresponds to overlapping orbits, and the single velocity-space loop traced by both vortices encloses a convex area~[Fig.~\ref{fig:polar_general}(a)]. 
In the rotating coordinate-space representation, the individual orbits are identically shaped ellipses as in Fig.~\ref{fig:2v_shared_rot_com}(b). The change induced in the dynamics when the initial point-vortex angular momentum $L$ is gradually increased is illustrated in Fig.~\ref{fig:polar_progression}. As $L$ is increased, the closed velocity-space orbit deforms and becomes concave, with the single minimum in the angular velocity splitting into two minima, each with the same angular velocity and opposite radial velocities [Fig.~\ref{fig:polar_progression}(a)]. In the rotating coordinate space, this corresponds to the development of a sharp point in the vortex paths near the edge of the trap, deforming the ellipses into droplets with their tips pointing away from the trap center. On further increasing $L$, this sharp point develops into a second loop in the path, creating a figure-eight curve in the rotating-frame coordinate space. In the velocity-space representation, the figure-eight stage corresponds to concave closed curves of the type shown in Fig.~\ref{fig:polar_progression}(b). Eventually a critical value of $L$ is reached at which the single loop in the velocity space self-intersects at zero radial velocity and a finite value of angular velocity, and subsequently separates into two nonintersecting simple loops [Fig.~\ref{fig:polar_progression}(c)]. Depending on the values of $L$ and $\phi$, one of the separated loops may lie inside the other [Fig.~\ref{fig:polar_progression}(c)], or they may not enclose any points in common [Fig.~\ref{fig:polar_progression}(d)].

\begin{figure}[tb]
\includegraphics[width=\columnwidth,keepaspectratio]{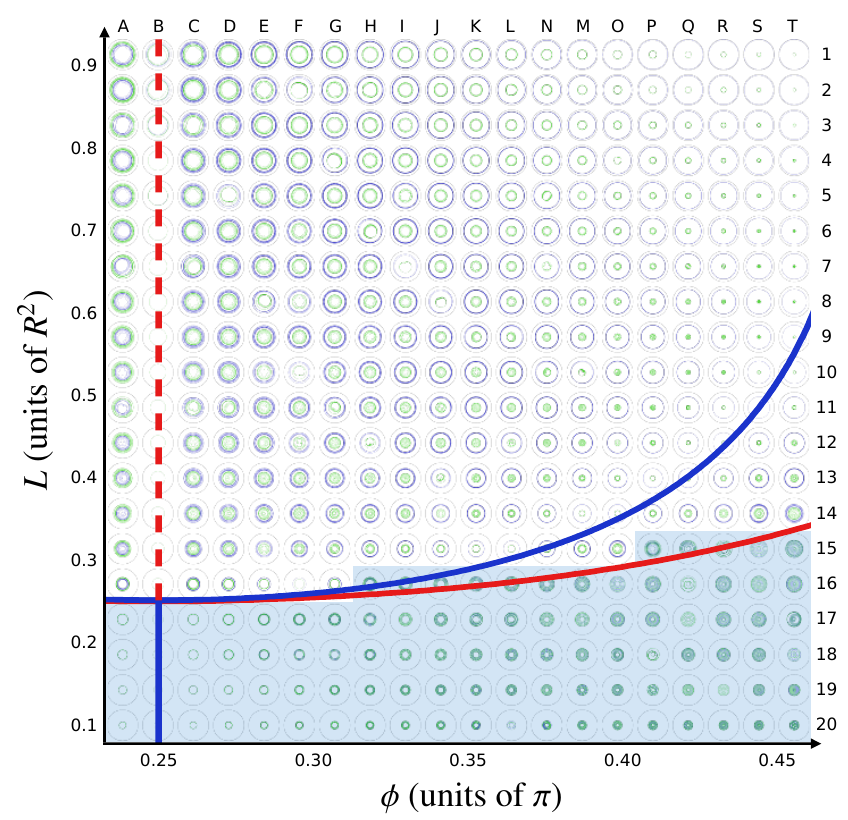}
\caption{\label{fig:phase_transitions} Laboratory-frame views of two-vortex dynamics, positioned according to their initial conditions in the 2D parameter space $\left(\phi,L\right)$; here the initial angle between the vortices is set to $\relangle=\pi$. The blue solid vertical line represents stable symmetric rigidly rotating states and the red dashed vertical line represents unstable symmetric rigidly rotating states. The blue (upper) curve represents the asymmetric rigidly rotating states, while the red (lower) curve denotes the topological transition between shared and separated phase spaces. In the shaded region below the red curve, the two vortices exhibit shared phase spaces. This figure should be compared with Fig.~1(c) in Ref.~\cite{Navarro2013a}. A high-resolution version of the diagram is provided in the Supplemental Material~\cite{Supplement}.} 
\end{figure}

Figure~\ref{fig:phase_transitions} demarcates the different types of two-vortex dynamics in the parameter space $\left(\phi,L\right)$ of different initial configurations with $\relangle=\pi$. Sampling of this $\left(\phi,L\right)$ space was done by scanning the parameters on a 20-by-20 grid of 400 initial conditions and integrating the system over a time interval of $60\Omega_0^{-1}$. Each grid cell in Fig.~\ref{fig:phase_transitions} shows the resulting dynamics in the position-coordinate space. A zoomable high-resolution version of this diagram is provided in the Supplemental Material~\cite{Supplement}.
The transition from a shared velocity-space loop (shaded region in Fig.~\ref{fig:phase_transitions}) to separated loops---i.e., from $\interval_1=\interval_2$ to $\interval_1\cap\interval_2=\emptyset$---is represented by the dark (red) solid line in Fig.~\ref{fig:phase_transitions}. For fixed $\phi=\phi_0$, values of $L$ above this transition point always result in distinct, nonintersecting orbits in both the coordinate-space [Figs.~\ref{fig:rigid_rot_stable}(b),~\ref{fig:general_motion_pos}(b), and~\ref{fig:2v_split_rot_com}] and the velocity-space representation [Figs.~\ref{fig:polar_general}(b),~\ref{fig:polar_progression}(c), and~\ref{fig:polar_progression}(d)]. This critical value of $L$ increases slightly with increasing $\phi$. 

The asymmetric rigidly rotating configurations are also indicated in Fig.~\ref{fig:phase_transitions} (upper, blue solid curve). We note in particular that these configurations lie inside the region of separated-phase-space dynamics and do not occur at the transition point between the two types except at a single point $\left(\phi,L,\relangle \right)=\left(\pi/4,\Lcrit,\pi\right)$. The symmetric rigidly rotating states, and the critical value $\Lcrit$ of the point-vortex angular momentum at which the bifurcation occurs along the line $\left(\phi,\relangle \right)=\left(\pi/4,\pi\right)$, are in agreement with previous predictions~\cite{Navarro2013a}. However, the topological change in the accessible phase space, where the vortex orbits become nonintersecting, was not reported in Ref.~\cite{Navarro2013a}. 

\begin{figure}[tb]
\includegraphics[width=\columnwidth,keepaspectratio]{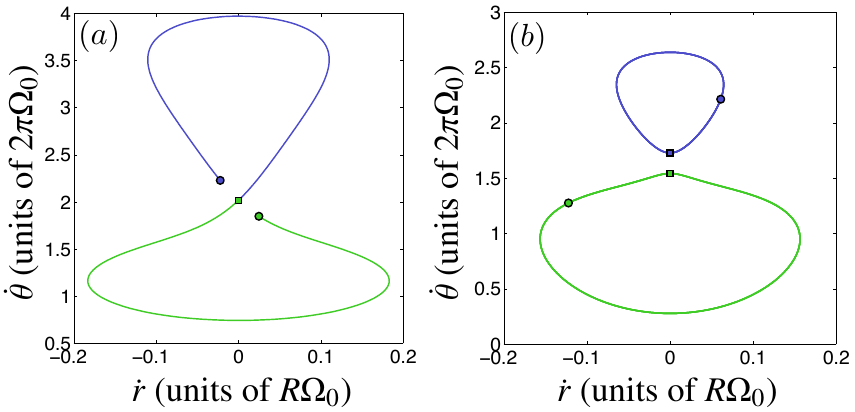}
\caption{\label{fig:polar_bifurcation} Comparison of (a) an unstable symmetric rigidly rotating state, with the initial conditions $\phi / \pi = 0.25$, $L = 0.955R^2$, and $\relangle=\pi$ (panel B1 in Fig.~\ref{fig:phase_transitions}), and (b) an asymmetric nonrigidly rotating state, with $\phi / \pi = 0.274$, $L = 0.64R^2$, and $\relangle=\pi$ [Fig.~\ref{fig:general_motion_pos}(b) and panel D8 in Fig.~\ref{fig:phase_transitions}].} 
\end{figure}

The change from stable rigidly rotating states to nonrigidly rotating ones can be understood by closely examining the symmetric rigidly rotating state that destabilizes at the bifurcation point $\Lcrit$. The unstable symmetric state with $L=0.955\,R^2 > \Lcrit$ (Fig.~\ref{fig:symmetric_rigid_unstable}) yields the simulated dynamics shown in Fig.~\ref{fig:polar_bifurcation}(a) using the velocity-space representation. Qualitatively, it resembles the velocity-space representations of states with $\phi \gtrsim \pi/4$ and  $L \gtrsim \Lcrit$ but in the latter the orbits of the two vortices in the velocity space become separated as shown in Fig.~\ref{fig:polar_bifurcation}(b).

Near the phase-space transition, the two-vortex system may, during its dynamics, approach the unstable rigidly rotating configuration but is then pushed away from it by the instability of the configuration.
If the dynamics exhibits shared phase spaces, this results in a swapping of the outside and inside vortices. For separated phase spaces, the radial velocity of each vortex changes sign, and the outside and inside vortices are pushed back into their respective zones. This suggests that at the bifurcation point $\left(\phi,L,\relangle \right)=\left(\pi/4,\Lcrit,\pi\right)$, the symmetric rigidly rotating states separate into two antisymmetric branches of asymmetric rigidly rotating states (which belong to the type of noncrossing orbits) and two symmetric branches of states on the phase-space-dividing boundary.
The branches of asymmetric rigidly rotating states are antisymmetric in the sense that the dynamics of the two rigidly rotating states with $\phi=\pi/4\pm\delta$ map to each by interchanging the two vortices. The phase-space-dividing branches are symmetric in the sense that, for values of $L$ at and below the branches, the initial states with $\phi=\pi/4\pm\delta$ and same $L$ represent essentially the same dynamics (due to time-translation and rotational symmetry of the model).

When the two vortices are not initially located at opposite sides of the trap, i.e., when $\relangle\neq\pi$, the rigidly rotating states become entirely absent but the phase-space separation transition persists. This is illustrated in Fig.~\ref{fig:schematic}(a): The symmetric and asymmetric rigidly rotating states form a bifurcating curve in the 2D plane $\relangle=\pi$ of the 3D parameter space $\left(\phi,L,\relangle\right)$ of possible two-vortex configurations. The phase-space-dividing boundary, on the other hand, constitutes a 2D surface. In the $\left(\phi,L,\relangle\right)$ space, all possible two-vortex orbits are planar curves (which may be single points) that are perpendicular to the $L$ axis (since $L$ is conserved) and never penetrate the phase-space wall.

\section{Discussion}\label{sec:discussion}

In conclusion, we have numerically studied the dynamics of two same-sign point vortices in a harmonically trapped superfluid. We discovered an impenetrable wall in the 3D phase space of possible two-vortex configurations that divides the ensuing vortex dynamics into two distinct types. In the first type, the two vortices move inside the same annular regions in the trap, whereas in the second type their orbits never intersect. The two types are particularly easy to distinguish in the 2D parameter space spanned by the angular and radial velocities of the vortices, where the first type results in one closed curve along which both vortices travel and the second type yields separate loops for each vortex. This phase-space wall is distinct from the bifurcation of rigidly rotating two-vortex configurations found by Navarro \emph{et al.}~\cite{Navarro2013a}. Importantly, the phase-space wall also persists for configurations where the two vortices are not initially at opposite sides of the trap center, unlike the rigidly rotating states.

Introducing the velocity-space representation opens a number of ways to extend the investigations of point-vortex dynamics in future studies. One obvious question is how the introduction of asymmetry between the vortices, i.e., $s_1\neq s_2$, would affect the transition phenomena in the phase space; the archetypal example of such a configuration is the vortex--antivortex pair ($s_2=-s_1$), which is known to exhibit stationary solutions in the harmonically trapped system~\cite{Freilich2010a,Kuopanportti2011a,Middelkamp2011a,Torres2011a}. On the other hand, increasing the number of vortices to three in Eqs.~\eqref{eq:PVM} results in the emergence of chaotic vortex dynamics in a particularly simple yet experimentally relevant setup; in the absence of the trap [i.e., for $\Omega_0=0$ in Eqs.~\eqref{eq:PVM}], chaos can reign only if $N\geq 4$. In fact, already the two-vortex case is likely to exhibit \emph{chaotic advection}~\cite{Are2002.PhysFluids14.1315}: If one formally introduces a third vortex with $s_3=0$, its motion in the flow field of the two genuine vortices may be chaotic. Chaotic advection is known to exist in the presence of three genuine vortices for $\Omega_0=0$~\cite{Aref1983a}.

Ultimately, the point-vortex model will serve as an efficient model for 2D quantum turbulence that corresponds to highly chaotic motion of a large number of point vortices. As such, it shows promise in further elucidating such phenomena as the inverse energy cascade, the emergence of Onsager vortices, and negative absolute Boltzmann temperatures associated with 2D turbulence in superfluids.   

\begin{acknowledgments}
We acknowledge financial support from the Australian Postgraduate Award (A.J.G.), the Emil Aaltonen Foundation (P.K.), the Finnish Cultural Foundation (P.K.), the Magnus Ehrnrooth Foundation (P.K.), the Technology Industries of Finland Centennial Foundation (P.K.), and the Australian Research Council via Discovery Project No.~DP130102321~(T.S.). We thank P.~J.~Easter for helpful comments on the manuscript.
\end{acknowledgments}
\bibliography{murray_two-vortex_dynamics.bbl}
\end{document}